\documentclass[aps,preprint,tightenlines,showkeys,nofootinbib]{revtex4}
\usepackage{amsmath,amssymb,graphicx}

\newcommand{\ND}{N^\dagger}
\newcommand{\VS}{\vec{\sigma}}
\newcommand{\VT}{\vec{\tau}}
\newcommand{\LRD}{\stackrel{\leftrightarrow}{\nabla}}
\newcommand{\aR}{{\cal C}^{(^3 \! S_1-^1 \! P_1)}}
\newcommand{\bR}{{\cal C}^{(^1 \! S_0-^3 \! P_0)}_{(\Delta I=0)}}
\newcommand{\cR}{{\cal C}^{(^1 \! S_0-^3 \! P_0)}_{(\Delta I=1)}}
\newcommand{\dR}{{\cal C}^{(^1 \! S_0-^3 \! P_0)}_{(\Delta I=2)}}
\newcommand{\eR}{{\cal C}^{(^3 \! S_1-^3 \! P_1)}}
\newcommand{\eftnopi}{\mbox{EFT($\not \! \pi$)}}
\newcommand{\CG}{{\cal G}}

\begin{document}

\title{An effective-field-theory analysis of low-energy parity-violation in nucleon-nucleon scattering }

\author {Daniel R. Phillips$^{1,3}$} \email{phillips@phy.ohiou.edu}
\author{Matthias R. Schindler$^1$}\email{schindle@ohio.edu}
\author{Roxanne P. Springer$^{1,2}$}\email{rps@phy.duke.edu}
\affiliation{$^1$Department of Physics and Astronomy, Ohio University, Athens, OH
45701, United States;\\
$^2$Department of Physics, Box 90305, Duke University, Durham, NC, 27708, United States\\
$^3$School of Physics and Astronomy, University of Manchester, Manchester, M13 9PL, United Kingdom}

\date{\today}

\begin{abstract}
We analyze parity-violating nucleon-nucleon scattering at energies $E < m_\pi^2/M$ using the effective field theory appropriate for this regime.
The minimal Lagrangian for short-range parity-violating  $NN$ interactions is written in an operator basis that encodes the five partial-wave transitions that dominate at these energies.  We calculate the leading-order relationships between parity-violating $NN$ asymmetries and the coefficients in the Lagrangian and also discuss the size of sub-leading corrections. We conclude with a discussion of further observables needed to completely determine the leading-order Lagrangian.
\end{abstract}

\maketitle

\section{Introduction}

The existence of parity violation in nuclear forces is a manifestation of the presence of weak interactions between the quarks in the nucleon. 
In this paper we discuss the most basic observables that display this phenomenon: $NN$ scattering asymmetries that would be zero were parity
conserved in the $NN$ interaction. We do this using an effective field theory (EFT) that is based on the existence of large ($\gg 1/m_\pi$) scattering 
lengths in the $NN$ system~\cite{We91,vanKolck:1997,KSW98A,KSW98B,Kaplan:1996xu,Gegelia:1998gn,BRMcG99,Chen:1999,review1,review2}. 
The scattering experiments we will study allow us to use an EFT that
contains only nucleon degrees of freedom, and treats all interactions between those degrees of freedom as being short-ranged. 
We choose to consider only the scattering length anomalously
large;  higher order corrections are obtained as a systematic
expansion in powers of $r/a$, where $a$ is the two-body scattering length and $r$ the effective range in the corresponding partial wave.

We find it convenient to examine such corrections to parity-violating $NN$ observables using a dynamical dibaryon field~\cite{Kaplan96,BG99,BS00}. This treatment captures the dominant dynamics in the situation where both the scattering length and effective range are unnaturally large. It corresponds to resumming an infinite subset of terms that are higher order in the power counting we use, where $r \sim 1/m_\pi \ll a \sim 1/p$. Therefore, expanding the results obtained from the dibaryon formalism in powers of $r/a$ reproduces the case of interest to us here, and so we employ dibaryon fields as a calculational tool.

There are a number of theoretical treatments based on hadronic degrees of freedom that have been used to study parity violation (PV). Pioneering theoretical studies 
on PV in nuclear forces were carried out by Danilov~\cite{Danilov}
 and Desplanques and Missimer~\cite{DesplanquesMissimer}. 
For most of the last thirty years, the framework of single-meson exchange, most commonly using the taxonomy developed by 
Desplanques, Donoghue, and Holstein (DDH) \cite{Desplanques:1979hn}, has been the one used to interpret and motivate experiments~\cite{Adelberger:1985ik,Haxton:2008ci,Page:1987ak}.  But the lack of a
concordance region in the space of DDH PV parameters (see, e.g., the plot in
Ref.~\cite{Haxton:2008ci}) may be related to the model assumptions---such as the mediation
of the PV $NN$ interaction by vector mesons---in that approach. The 
consequent desire by the community to ``...recast the DDH language....in
terms of...effective short-range parity-violating N-N
interactions"~\cite{ECT2000} has motivated EFT treatments of the PV $NN$
force. 

Recently the version of chiral perturbation theory appropriate for few-nucleon systems~\cite{We91,ORvK96,Ep99,EM01,Ep05,Phillips07} ($\chi$ET hereafter) has been
used to derive the long-range ($r \sim 1/m_\pi$) part of the PV $NN$ force, and to classify the short-distance operators that appear in that force~\cite{Zhu:2004vw}. (See also the 
original $\chi$PT analysis of PV pion-nucleon operators in Ref.~\cite{Kaplan:1992vj}). This has the advantage that---up to a given order in $\chi$ET---one can guarantee that a complete set of parity-violating operators has been considered. 
The potential of Ref.~\cite{Zhu:2004vw} is formulated in terms of the appropriate degrees of freedom for momenta of order $m_\pi$: nucleons and pions.\footnote{For extensions to include the $\Delta$ isobar see Refs.~\cite{Kaiser:2007zzb,Liu:2007fn}.} The use of heavier mesons to encode $NN$ interactions of different t-channel quantum numbers is not necessary at energies below $\sim$ 200 MeV. By fitting the constants that encode the short-distance $NN$ interaction to data, the $\chi$ET treatment of PV in $NN$ scattering avoids any assumption about what dynamics is at work for $r  \ll 1/m_\pi$. 
The consequences of $\chi$ET for PV $NN$ scattering, as well as other PV few-nucleon-system observables, have been computed in Refs.~\cite{Liu:2006dm,Hyun:2006cb,Schiavilla:2008}. However, the PV operators and strong-nuclear-force wave functions employed in these works were not consistent, since phenomenological models were used for the latter, but a $\chi$ET for the former. Refs.~\cite{Liu:2006dm,Schiavilla:2008} also include what are referred to as ``pionless-theory" results, and there the mismatch between operators and wave functions is a serious problem. The use of AV18 wave functions for one, and a pionless EFT for the other, involves a mismatch of roughly an order of magnitude in the resolution ($\sim$ renormalization) scales of these two different calculational ingredients.

The pionless theory, \eftnopi, is relevant to studies of parity violation because 
many of the existing and planned experiments ~\cite{Eversheim:1991tg,Kistryn:1987tq,Lauss:2006es,Knyazkov:1984ke,Cavaignac:1977uk,Earle:1988fc,Stiliaris:2005hh,JLAB} take place at energies at or below 10's of MeV. In this region the pion-exchange nature of the
nuclear force is not resolved. For energies $E < m_\pi^2/M$ the EFT in which interactions are encoded as $NN$ contact operators contains all
the relevant degrees of freedom.  The convergence of this EFT for two-nucleon-system observables is well demonstrated~\cite{review1,review2,Rupak:1999rk,Butler:2001jj} in the low-energy regime.
In contrast, there have been significant questions raised recently about the appropriate power counting for short-distance operators in the $NN$ system in the $\chi$ET where pions are explicit degrees of freedom~\cite{Kaplan:1996xu,Beane:2001bc,NTvK,EM06,PVRA,Birse07}, as well as about whether the Delta(1232) needs to be included as an explicit degree of freedom in order to guarantee reasonable convergence~\cite{ORvK96,Pa05,Ep07}.

In \eftnopi, parity violation in $NN$ scattering is described by contact operators that have a lowest possible dimension of seven. As Girlanda has recently shown~\cite{Girlanda:2008ts} there are five such independent operators.\footnote{While Ref.~\cite{Zhu:2004vw} lists ten operators, it is pointed out that
only five different combinations are relevant at low energies.}
These five leading-order EFT contact operators contain the same physics as the five Danilov amplitudes \cite{Danilov} that encode the mixing between S and P-waves that becomes possible in the presence of parity violation. In Section~\ref{sec-Lag} we rewrite the Lagrangian of Ref.~\cite{Girlanda:2008ts} in terms of five operators that each mediate a specific S-P transition. This makes the computation of longitudinal asymmetries in $NN$ scattering (given in Sec.~\ref{sec-results}) straightforward. We present analytic results for the longitudinal asymmetries in $nn$, $pp$, and $np$ scattering. We use a dynamical dibaryon field to obtain 
a portion of the higher-order corrections in $r/a$ and $rp$ in the \eftnopi~expansion for the strong rescattering. Subsection~\ref{sec-Coulomb} computes the Coulomb effects that are present in $pp$ scattering at low energy and so modify the 
expression in the $pp$ case.
We extract the scale dependence of the result obtained when assuming that the $r$ dependence is higher order, and then discuss the $r$ dependent corrections. In Sec.~\ref{sec-experiment} we compare our results with the two existing pieces of experimental data~\cite{Eversheim:1991tg,Kistryn:1987tq} that are within the range of validity of this EFT. We close in Sec.~\ref{sec-conclusion} with a summary and a discussion of further $NN$ system experiments that could pin down the LO PV \eftnopi~Lagrangian. Details of the conversion of one set of operators to another are given in an Appendix.

\section{Lagrangians}

\label{sec-Lag}

At low enough energies, the details of the gauge boson ($g$, $W^\pm$, $Z$)
exchange between interacting quarks in the two-nucleon system are not 
experimentally accessible.  Instead, the system can be described by
using nucleon interpolating fields and treating both strong and weak
interactions as contact interactions.  To a given order there are a finite
number of independent operators that describe these interactions.

The Lagrangian for the parity-conserving two-nucleon sector may be written as \cite{We91}
\begin{equation}\label{Lag:PC}
\mathcal{L} = \ND(i \partial_0 + \frac{\vec{\nabla}^2}{2M})N -\frac{1}{2}C_S(\ND N)^2-\frac{1}{2}C_T(\ND \VS N)^2 + \ldots,
\end{equation}
where the ellipsis stands for terms that contain more derivatives, and the nucleon field $N$ carries both isospin and spin indices. The $\sigma_i$ are
the SU(2) Pauli matrices in spin space
and the $\tau_a$ will be the SU(2) Pauli matrices in isospin space.  In Eq.~(\ref{Lag:PC}) the $NN$ effective ranges are assumed to be ``natural" compared to the expected scale of $1/m_\pi$ while the $NN$ scattering lengths are large in the same units. This facilitates an expansion of 
the $NN$ amplitude in powers of the small parameter $Q$, where $Q \sim 1/a \sim p$ (with $p$ the $NN$ relative 
momentum)~\cite{vanKolck:1997,KSW98A,Gegelia:1998gn}.

The same physics can be expressed using an operator basis that makes
the incoming and outgoing partial waves explicit \cite{Kaplan:1998sz,Savage:1998rx},
\begin{align}\label{Lag:PCpartial}
\mathcal{L} = & \ND(i \partial_0 + \frac{\vec{\nabla}^2}{2M})N -
\frac{1}{8}{\cal C}_0^{(^1 \! S_0)} (N^T \tau_2 \tau_a \sigma_2 N)^\dagger 
(N^T \tau_2 \tau_a \sigma_2 N) \   \notag\\
&- \frac{1}{8}{\cal C}_0^{(^3 \! S_1)} (N^T \tau_2 \sigma_2 \sigma_i N)^\dagger 
(N^T \tau_2  \sigma_2 \sigma_i N) + \ldots,
\end{align}
where now
${\cal C}_0^{(^1 \! S_0)}=C_S-3 C_T$ and ${\cal C}_0^{(^3 \! S_1)}=C_S+C_T$.
The operator between nucleons is simply
the projector onto the relevant partial wave, with the normalized projectors being \cite{Kaplan:1998sz,Savage:1998rx}
$$P_a(^1 \! S_0)={1 \over \sqrt{8}} \tau_2 \tau_a \sigma_2 \ ; \ \ \ \ \ 
P_i(^3 \! S_1)={1 \over \sqrt{8}} \tau_2 \sigma_2 \sigma_i  \ \ . $$

A convenient form of this Lagrangian is provided by use of dibaryon fields~\cite{Kaplan96,BG99,BS00}. This form is equivalent to Eqs.~(\ref{Lag:PCpartial}) and (\ref{Lag:PC}) at leading order, but resums all the higher-order corrections in $NN$ scattering that are proportional to the effective range. It would therefore give the exact $NN$ amplitude in both the ${}^1S_0$ and ${}^3S_1$ channel were the shape parameter and all higher-order terms in the effective-range expansion zero. Dibaryon fields, $s_a$ and $t_i$, respectively, for the $^1S_0$ and $^3S_1$ states, are included in the Lagrangian~\cite{BS00,BGHR}: 
\begin{eqnarray}
{\cal L}&=&N^\dagger \left(i \partial_0 + \frac{\vec{\nabla}^2}{2M}\right)N - t_i^\dagger \left(i \partial_0 + \frac{\vec{\nabla}^2}{4M} - \Delta_{(^3\!S_1)}\right) t_i - g_{(^3 \! S_1)} \left[t_i^\dagger N^T P_i(^3 \! S_1) N + \mbox{h.c.}\right]  \nonumber\\
&& \qquad - s_a^\dagger \left(i \partial_0 + \frac{\vec{\nabla}^2}{4M} - \Delta_{(^1\!S_0)}\right) s_a - g_{(^1 \! S_0)} \left[s_a^\dagger N^T P_a(^1 \! S_0) N + \mbox{h.c.}\right],
\label{Lag:dib}
\end{eqnarray}
with $a$ an isospin and $i$ a spin index. If PDS (power divergence subtraction)~\cite{KSW98A,KSW98B} is used to compute loops then for both channels we have (in an obvious notation with channel subscripts suppressed):
\begin{equation}
g^2=\frac{8 \pi}{M^2 r}; \quad \Delta=\frac{2}{M r}\left(\frac{1}{a} - \mu \right).
\end{equation}

The leading-order PV two-nucleon Lagrangian can also be expressed
in a variety of bases.  Paralleling the one used in Eq.~(\ref{Lag:PC}),
Ref.~\cite{Girlanda:2008ts} writes:
\begin{align}\label{Lag:Gir}
\mathcal{L}_{PV}^{Gir}=&\left\{\CG_1 (\ND \VS N \cdot \ND i \LRD N -\ND N \ND i \LRD\!\cdot\;\VS    N) \right. \notag\\
& -\tilde \CG_1 \epsilon_{ijk} \ND \sigma_i N \nabla_j(\ND \sigma_k N) \notag\\
& -\CG_2 \epsilon_{ijk}\left[ \ND \tau_3\sigma_i N \nabla_j(\ND \sigma_k N) + \ND \sigma_i N \nabla_j (\ND \tau_3\sigma_k N) \right] \notag\\
& -\tilde \CG_5 \mathcal{I}_{ab}\epsilon_{ijk}\ND \tau_a\sigma_i N \nabla_j(\ND \tau_b \sigma_k N) \notag\\
& + \left. \CG_6 \epsilon_{ab3}\vec{\nabla}(\ND \tau_a N)\cdot \ND \tau_b \VS N \right\},
\end{align}
where  $a\!\LRD\! b=a\vec{\nabla}b-a\!\stackrel{\leftarrow}{\nabla}\!b$ and
\[ \mathcal{I}=
\begin{pmatrix} 
1 & 0 & 0 \\
0 & 1 & 0\\
0 & 0 & -2
\end{pmatrix}. \]
Note that we have renamed the coefficients as compared to Ref.~\cite{Girlanda:2008ts}, in order to avoid confusion with the parity-conserving Lagrangian. In doing so we absorbed into the
$\CG$s the overall normalization factor of $1/\Lambda_\chi^3$ that multiplies the $C_i$'s and $\tilde{C}_i$ 's
in Ref.~\cite{Girlanda:2008ts}.  In \eftnopi~the coefficients are
typically dependent on the renormalization point, $\mu$, used in the evaluation of loop diagrams. 
Using the partial-wave basis, as in Eq.~(\ref{Lag:PCpartial}), we have:
\begin{align}\label{Lag:RPS}
\mathcal{L}_{PV}^{PW}= & - \left[ \aR \left(N^T\sigma_2 \ \VS
\tau_2 N \right )^\dagger \cdot  \left(N^T \sigma_2 i \LRD \tau_2 N\right) \right. \ + \notag\\
& \bR \left(N^T\sigma_2 \tau_2 \VT N\right)^\dagger  \left(N^T\sigma_2 
\ \VS \cdot i \LRD \tau_2 \VT N\right) \ + \notag\\
& \cR \ \epsilon^{3ab} \left(N^T\sigma_2 \tau_2 \tau^a N\right)^\dagger \left(N^T \sigma_2  \ \VS\cdot \LRD \tau_2 \tau^b N\right) \ + \notag\\
& \dR \ \mathcal{I}^{ab} \left(N^T\sigma_2 \tau_2 \tau^a N\right)^\dagger \left(N^T \sigma_2 \ \VS\cdot i \LRD \tau_2 \tau^b N\right) \ + \notag\\
& \left. \eR \ \epsilon^{ijk} \left(N^T\sigma_2 \sigma^i \tau_2 N\right)^\dagger  \left(N^T \sigma_2 \sigma^k \tau_2 \tau_3 \LRD^j N\right)\right] + h.c..
\end{align}

The two Lagrangians in Eq. (\ref{Lag:Gir}) and (\ref{Lag:RPS}) give the same results for physical observables if the low-energy constants obey the relationships (see Appendix):
\begin{align}\label{Lag:LECRel}
\aR&={1 \over 4} (\CG_1-\tilde \CG_1)  \ \ , \notag\\
\bR&={1 \over 4} (\CG_1+\tilde \CG_1) \ \ , \notag \\
\cR&={1 \over 2} \CG_2 \notag \ \ , \\
\dR&=-{1 \over 2} \tilde \CG_5 \notag  \ \ , \\
\eR&={1 \over 4} \CG_6 \ \ .
\end{align}
Note that there are
five independent coefficients at this order---as explained in Refs.~\cite{Zhu:2004vw,Holstein:2006bv,Girlanda:2008ts}.  They dictate the
only possible nucleon-nucleon scattering observables at low enough
(i.e., non-dynamical pion) energies.  From the partial-wave point of
view, only the coefficients $\aR$, $\bR$, $\dR$, and $\eR$
are
involved in parity-violating neutron-proton observables; while
$\bR$, $\cR$, and $\dR$ 
are involved in parity-violating neutron-neutron (or
proton-proton) observables.  

In the DDH approach, all but the $^3 \! S_1-^3 \! \! P_1$ operator are 
considered only in terms of vector-meson exchange.  At low energies, the
vector mesons are not dynamical, so the DDH description can
be considered a way to ``encode'' the processes so that
calculations and experiments can be compared, so long as
the vector-meson interpretation is not taken literally. In particular,
at very low energies only five independent parameters are relevant for
the physics of parity violating $NN$ scattering---they are sufficient
to encode all leading-order phenomena. The EFT parameterization presented here provides a model-independent language in which to compare experiments.

\section{Results}

\label{sec-results}

\subsection{Calculation of longitudinal analyzing power}

\begin{figure}[b]
\includegraphics{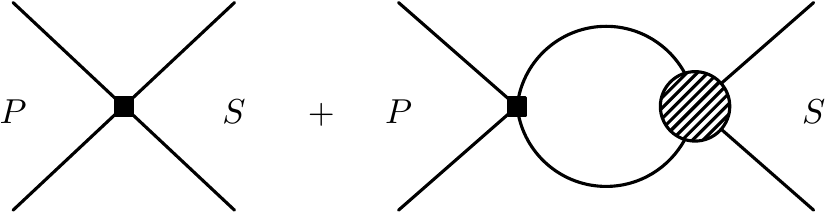}\caption{Diagrams contributing to parity-violating NN scattering. The shaded blob is the leading order parity-conserving amplitude, while the square denotes the vertex from the parity-violating Lagrangian\label{Fig:scatt}. The nucleons are in a P-wave on the left of the parity-violating vertex, and in an S-wave on the right.}
\end{figure}

Parity violation in the $NN$ interaction leads to mixing between odd and even partial waves. To obtain the leading effects of this mixing in \eftnopi~it is sufficient to calculate the amplitude that mediates $S$-wave to $P$-wave transitions. The mixing of higher partial waves is suppressed by additional powers of the small parameter $Q$.

The leading diagrams contributing to this parity-violating $NN$ scattering amplitude are shown in Fig.~\ref{Fig:scatt}. Note that only the $S$-wave side receives an enhancement from the strong $S$-wave bubble sum.  Diagrams with strong rescattering on the $P$-wave side are higher order.  We evaluate the diagrams shown in Fig.~\ref{Fig:scatt} using the PDS~\cite{KSW98A,KSW98B} renormalization scheme to calculate the loops. Keeping in mind the issue of higher-order corrections (see Subsec.~\ref{sec-reneq0}), we employ the dibaryon Lagrangian (\ref{Lag:dib}) to compute the strong rescattering. Our result for the scattering amplitude is:
\begin{equation}
T_{nn/pp}^{PV} = \pm 4 \, p \, \, {\cal A}_{nn/pp} \left(\frac{1}{a^{^1 \! S_0}}-\frac{1}{2}r^{^1 \! S_0}p^2 - \mu \right) \, \, \left(\frac{1}{a^{^1 \! S_0}}-\frac{1}{2}r^{^1 \! S_0}p^2+ip\right)^{-1} ,
\label{eq:nnampl}
\end{equation}
and
\begin{eqnarray}
T_{np}^{PV} &=& \pm 4 \,p \, \,\left[{\cal A}_{np}^{{}^1 \! S_0} \left(\frac{1}{a^{^ 1\!S_0}}-\frac{1}{2}r^{^1\!S_0}p^2 - \mu \right) \, \, \left( \frac{1}{a^{^1\!S_0}}-\frac{1}{2}r^{^1\!S_0}p^2+ip \right)^{-1} \right.\nonumber\\
&&\left. + {\cal A}_{np}^{{}^3\!S_1} \left(\frac{1}{a^{^3\!S_1}}-\frac{1}{2}r^{^3\!S_1}p^2 - \mu\right) \, \, \left(\frac{1}{a^{^3\!S_1}}-\frac{1}{2}r^{^3\!S_1}p^2+ip\right)^{-1}\right],
\label{eq:npampl}
\end{eqnarray}
where $-iT$ is the sum of diagrams in Fig.~\ref{Fig:scatt}, $p=|\vec p|$ (see Fig.~\ref{Fig:match}) and the upper (lower) sign is for a beam of positive (negative) helicity.
The (strong) parameters $a^{^{2S+1}L_J}$  and $r^{^{2S+1}L_J}$ are, respectively, the scattering length and effective range of a particular partial wave.
The weak $NN$ interaction parameters are collected in amplitudes ${\cal A}$, which are given by:
\begin{align}
{\cal A}_{nn} &= \CG_1 +\tilde \CG_1 -2(\CG_2+\tilde \CG_5) \label{eq:A1}\\
&= 4\left( \bR -  \cR +  \dR \right), \label{eq:A1b}\\
{\cal A}_{pp} &= \CG_1 +\tilde \CG_1 +2(\CG_2-\tilde \CG_5)\\
&= 4 \left( \bR + \cR + \dR \right), \label{eq:A2}\\
{\cal A}_{np}^{^1S_0} &= \CG_1 +\tilde \CG_1 +4 \CG_5\\
&= 4 \left(\bR - 2 \dR \right), \label{eq:A3}\\
{\cal A}_{np}^{^3S_1} &= \CG_1 -\tilde \CG_1 -2 \CG_6\\
&= 4 \left(\aR -2 \eR \right) \label{eq:A4},
\end{align}
in the notation of the Lagrangians (\ref{Lag:Gir}) and (\ref{Lag:RPS}), respectively. While the $T^{PV}_{NN}$ expressions appear to have an explicit subtraction point ($\mu$) dependence, as physical observables each must be $\mu$ independent. This dictates the scaling of the ${\cal A}_{NN}$ with respect to $\mu$.

The leading-order  [$O(Q^0)$]  amplitudes are obtained by setting the effective ranges $r^{^1\!S_0}$ and $r^{^3\!S_1}$ equal to zero in Eqs.~(\ref{eq:nnampl}) and (\ref{eq:npampl}). This yields, for example,
\begin{equation}
T^{PV}_{nn}(r^{^1\!S_0}=0)=\pm 4p \, \frac{{\cal A}_{nn}}{{\cal C}_0^{(^1 \! S_0)}}  \, \frac{4 \pi}{M} \frac{1}{\frac{1}{a^{{}^1\!S_0}}+ i p},
\label{eq:LOnn}
\end{equation}
a result that already appeared in Section 4 of  Ref.~\cite{Zhu:2004vw}, but here we have also provided the relationship of the parity-violating amplitude ${\cal A}_{nn}$ to the coefficients in the Lagrangian(s) (Eqs.~(\ref{eq:A1}) and (\ref{eq:A1b})). The ratio of ${\cal A}_{nn}$ to ${\cal C}_0^{(^1 \! S_0)}$ must be independent of $\mu$. Since, in PDS,
\begin{equation}
{\cal C}_0^{(^1 \! S_0)}=\frac{4 \pi}{M} \frac{1}{\frac{1}{a^{^1 \! S_0}} - \mu},
\label{eq:C0}
\end{equation}
we have
\begin{equation}
\frac{1}{{\cal A}_{nn}} \frac{\partial {\cal A}_{nn}}{\partial \mu}=\frac{1}{\frac{1}{a^{^1 \! S_0}} - \mu}.
\end{equation}

In $NN$ scattering at these energies the weak interaction is about $10^{-7}$ times the strong interaction. Therefore
feasible experiments involve observables that
vanish under strong interactions.  Relevant $NN$ measurements have focused on longitudinal asymmetries in $\vec{N} + N$ scattering.
Here, the interference terms between the strong and weak operators
change sign when the longitudinal polarization of the incoming nucleon changes
sign.  The strong-interaction scattering is unaffected by a change
in polarization, so an asymmetry is formed when the differential
cross sections of the two different polarization states are subtracted.

From the scattering amplitude calculated above we can determine the longitudinal asymmetry:
\begin{equation}\label{eq:ALdef}
A_L=\frac{\sigma_+ - \sigma_-}{\sigma_+ + \sigma_-},
\end{equation} 
where $\sigma_\pm$ is the total scattering cross section of a nucleon with helicity $\pm$ on an unpolarized nucleon target---unless integration over a restricted angular range (e.g. in $pp$ scattering) is indicated. Neglecting, for the moment, the Coulomb interaction in the $pp$ case we find
\begin{equation}
A_L^{nn}=\frac{2M}{\pi}p\, {\cal A}_{nn} \left(\frac{1}{a^{^1\!S_0}}-\frac{1}{2}r^{^1\!S_0}p^2-\mu \right),
\label{eq:ALnn}
\end{equation}
\begin{equation}
A_L^{pp}=\frac{2M}{\pi}p\, {\cal A}_{pp} \left(\frac{1}{a^{^1\!S_0}}-\frac{1}{2}r^{^1\!S_0}p^2-\mu \right),
\label{eq:ALpp}
\end{equation}
and
\begin{align}
A_L^{np}= &\frac{2M}{\pi}p \left\{ \frac{ \frac{d\sigma^{^1\!S_0}}{d\Omega}}{\frac{d\sigma^{^1\!S_0}}{d\Omega}+3\frac{d\sigma^{^3\!S_1}}{d\Omega}}\,{\cal A}_{np}^{^1\!S_0}\left(\frac{1}{a^{^1\!S_0}}-\frac{1}{2}r^{^1\!S_0}p^2-\mu \right) \right. \notag\\
& \left.  + \frac{ \frac{d\sigma^{^3\!S_1}}{d\Omega}}{\frac{d\sigma^{^1\!S_0}}{d\Omega}+3\frac{d\sigma^{^3\!S_1}}{d\Omega}} \, {\cal A}_{np}^{^3\!S_1}\left(\frac{1}{a^{^3\!S_1}}-\frac{1}{2}r^{^3\!S_1}p^2-\mu \right) \right\}.
\label{eq:ALnp}
\end{align}
The differential cross sections $\frac{d\sigma^{^1\!S_0}}{d\Omega}$ and $\frac{d\sigma^{^3\!S_1}}{d\Omega}$ only contain contributions from the parity-conserving Lagrangian (see Eqs.~(\ref{Lag:PC}) and (\ref{Lag:PCpartial})):
\begin{equation}
\frac{d \sigma}{d \Omega}=\left[\left(\frac{1}{a} - \frac{1}{2} r p^2\right)^2 + p^2\right]^{-1}.
\end{equation}
We have again suppressed the channels' superscripts.

Upon setting $r^{{}^1\!S_0}=0$ and using Eq.~(\ref{eq:C0}), Eqs.~(\ref{eq:ALnn}) and (\ref{eq:ALpp}) recapture the form derived in Refs.~\cite{Zhu:2004vw,Holstein:2006bv}:
\begin{equation}
A_L^{pp/nn}=8 p \frac{{\cal A}_{pp/nn}}{{\cal C}_0^{^1\!S_0}} \ \ .
\label{eq:ALppLO}
\end{equation}
This, and the more complex formula for $A_L^{np}$:
\begin{equation}
A_L^{np}= 8 p \left( \frac{ \frac{d\sigma^{^1\!S_0}}{d\Omega}}{\frac{d\sigma^{^1\!S_0}}{d\Omega}+3\frac{d\sigma^{^3\!S_1}}{d\Omega}}\,\frac{{\cal A}_{np}^{^1\!S_0}}{{\cal C}_0^{^1\!S_0}} + \frac{ \frac{d\sigma^{^3\!S_1}}{d\Omega}}{\frac{d\sigma^{^1\!S_0}}{d\Omega}+3\frac{d\sigma^{^3\!S_1}}{d\Omega}} \, \frac{{\cal A}_{np}^{^3\!S_1}}{{\cal C}_0^{^3\!S_1}}\right),
 \end{equation}
are the LO predictions of \eftnopi~for these asymmetries. The only unknown quantities in these predictions are the coefficients of the parity-violating Lagrangian. Eqs.~(\ref{eq:ALnn})--(\ref{eq:ALnp}) relate these coefficients (Eqs.~(\ref{eq:A1})--(\ref{eq:A4})) to observable asymmetries. From these expressions we see that a measurement of all three analyzing powers as a function of energy could pin down {\it four} different combinations of coefficients, since the two pre-factors in Eq.~(\ref{eq:ALnp}) have distinct energy dependence---even if $r^{^1\!S_0}=r^{^3\!S_1}=0$. However, the $nn$ experiment is not feasible in the foreseeable future, and so alternative strategies to access the combination ${\cal A}_{nn}$ are probably necessary.

\subsection{Coulomb corrections for $pp$}

\label{sec-Coulomb}

The result in Eq. (\ref{eq:ALppLO}) ignores the Coulomb interaction. Coulomb photons can be included in \eftnopi, and the computation of Coulomb scattering was carried out to leading order for S-wave $NN$ scattering in Ref.~\cite{Kong:1998sx,Kong:1999sf}.

In the parity-violating case the computation proceeds as in Fig.~\ref{Fig:scatt}, except now Coulomb photons must be added to the initial, final, and all intermediate states. Since the initial-state and final-state Coulomb scattering factorizes this  yields the final result, quoted in Ref.~\cite{Zhu:2004vw,Holstein:2006bv}:
\begin{equation}
T_{NC}^{PV} (r^{^1\!S_0}=0)=\pm4 p \frac{{\cal A}_{pp}}{{\cal C}^{^1\!S_0}_0} C_\eta^2 \exp(i(\sigma_0(\eta) + \sigma_1(\eta)))\frac{1}{\frac{1}{C^{^1\!S_0}_0} - J_0(p)},
\end{equation}
where the purely Coulombic part of the scattering amplitude has been separated off the total amplitude \cite{GoldbergerWatson},
\begin{equation}
T=T_{NC} + T_{Coul}.
\label{eq:sum}
\end{equation}
$C_\eta^2$ is the Sommerfeld factor:
\begin{equation}
C_\eta^2=\frac{2 \pi \eta}{e^{2 \pi \eta}-1},
\end{equation}
with the Coulomb parameter $\eta \equiv \frac{M \alpha}{2 p}$, and $\sigma_l(\eta)={\rm arg} \Gamma(l + 1 + i \eta)$, where $\Gamma$ is the Euler gamma function.
$J_0(p)$ is the Coulomb-modified bubble:
\begin{equation}
J_0^{\rm finite}(p)=-\frac{\alpha M^2}{4 \pi}\left[H(\eta) - \ln\left(\frac{\mu \sqrt{\pi}}{\alpha M}\right) - 1 + \frac{3}{2} C_E\right] - \frac{\mu M}{4 \pi}
\label{eq:J0finite}
\end{equation}
once divergences in $D=4$ and $D=3$ have been subtracted, 
\begin{equation}
H(\eta)=\psi(i\eta) + \frac{1}{2 i \eta} - \log(i\eta),
\end{equation}
with $\psi$ the derivative of the Euler Gamma function, and $C_E=0.5772 (...) $ is Euler's constant. (See also Refs.~\cite{Barford:2002je,Ando:2008}.)

Experimental asymmetries are typically measured over a finite angular range. This is a particularly important detail in $pp$ scattering, due to the infinite Coulomb cross section in the forward direction. Implementing the integrals over a finite range $\theta_1 \leq \theta \leq \theta_2$ we have:
\begin{eqnarray}
A_L^{pp}&=&\frac{\int_{\theta_1}^{\theta_2} d \theta \, \sin \theta \, \, 2 \, {\rm Re}[(T_{NC}^{PV}) (T_{NC} + T_{Coul})^\dagger]}{\int_{\theta_1}^{\theta_2} d \theta \, \sin \theta \, |T_{NC} + T_{Coul}|^2} \nonumber \\
&\approx&2  \, {\rm Re}\left[ \frac{T^{PV}_{NC}}{T_{NC}} \left({1 - \frac{1}{\cos \theta_1 - \cos \theta_2} \int_{\theta_1}^{\theta_2} d  \theta \sin \theta \, \, \frac{T_{Coul}}{T_{NC}}}\right) \right] \ \ ,
\end{eqnarray}
where we have used the fact that $T_{NC,PV}$ and $T_{NC}$ are angle independent 
at this order, and have neglected $|T_{Coul}|^2$.
Ref.~\cite{Kong:1998sx} finds that $$T_{NC}=
{C_\eta^2 \ e^{2 i \sigma_0(\eta)} \over { 1 \over {\cal C}_0^{^1 \! S_0}} - J_0(p)} \ \ ,$$
yielding
\begin{eqnarray}
A_L^{pp}&\approx& 8 p \frac{{\cal A}_{pp}}{{\cal C}^{^1\!S_0}_0} {\rm Re} \left[ e^{i[\sigma_1(\eta) - \sigma_0(\eta)]}\left({1 - \frac{1}{\cos \theta_1 - \cos \theta_2} \int_{\theta_1}^{\theta_2} d  \theta \sin \theta \, \, \frac{T_{Coul}}{T_{NC}}}\right)\right].
\label{eq:ratios}
\end{eqnarray}
The factor in square brackets contains the Coulomb corrections to the result of the previous section. It is a function of $\eta$, the scattering length $a$, and the angular range being examined.

For the experiments of interest here we have $\eta \ll 1$. For small $\eta$,
\begin{equation}
T_{Coul}=\frac{2 \pi \alpha}{p^2} \frac{1}{1- \cos \theta} + O(\alpha^2),
\end{equation}
and
\begin{equation}
T_{NC}=\frac{4 \pi}{M}\frac{1}{\frac{1}{a_{S}(\mu)} + ip} + O(\eta),
\end{equation}
with the strong $pp$ scattering length, $a_S(\mu)$, defined by: 
\begin{equation}
\frac{4 \pi}{M {\cal C}^{^1S_0}_0(\mu)}=\frac{1}{a_S(\mu)} - \mu.
\end{equation}
A reliable extraction of $a_S(\mu)$ from $pp$ data appears to require a computation to several orders in \eftnopi~\cite{Ando:2007}. Instead,  for comparison to experiments in Sec.~\ref{sec-experiment}, we use isospin symmetry, and take for $a_S(\mu)$ the `recommended' central value of the $nn$ scattering length, $a_{nn}=-18.59$ fm~\cite{MS2001}.
We obtain
\begin{equation}
A_L^{pp}\approx 8 p {{\cal A}_{pp} \over {\cal C}^{^1\!S_0}_0} {\rm Re} \ 
\left[e^{i[\sigma_1(\eta) - \sigma_0(\eta)]} \left\{1 + \eta \left(\frac{1}{a_S(\mu) p} + i\right) \frac{1}{\cos \theta_1 - \cos \theta_2} \ln\left(\frac{1- \cos \theta_1}{1- \cos \theta_2}\right)\right\}\right]
\end{equation}
\begin{equation}
= 8 p \frac{{\cal A}_{pp}}{{\cal C}^{^1S_0}_0}\left[1 + \eta \left(\frac{1}{a_S(\mu) p} \right) \frac{1}{\cos \theta_1 - \cos \theta_2} \ln\left(\frac{1- \cos \theta_1}{1- \cos \theta_2}\right) + {\cal O}(\eta)^2 \right] \ \ ,
\label{eq:ALppfinal}
\end{equation}
so long as forward angles are avoided.

Even for $pp$ experiments at $T_{\rm lab}=0.5$ MeV, we have $\eta \approx 0.22$, so $\eta$ should be a good expansion parameter. Since parity-violating asymmetries grow as $\sqrt{T_{\rm lab}}$ the extant measurements of $A_L$ were conducted at energies significantly higher than this, so in practice ignoring Coulomb (as was done in the pioneering study of Ref.~\cite{DesplanquesMissimer}), or expanding in powers of $\eta$,
is a good approximation. 

This suggests using a different expansion where effects proportional to $M \alpha/p$ are treated perturbatively. 
However,   numerically $M \alpha \sim 1/a$ in the ${}^1S_0$ channel. If $M \alpha/p$ is treated as a small parameter, then $1/(ap)$ should really also be treated as a perturbation. This results in a theory set up as an expansion around the unitary ($|a| \rightarrow \infty$) limit~\cite{HammerHiga:2008}. Attempts to treat Coulomb interactions in perturbation theory and retain the $1/(ap)$ corrections in the unitary limit to all orders requires care since the divergences that are present in the Coulomb bubble must still be absorbed~\cite{Kong:1998sx,Gegelia:2003ta}. While the Coulomb contributions to the final result, Eq.(\ref{eq:ALppfinal}), are small for all existing and proposed experiments, here we have retained them to all orders in the intermediate steps of the calculation of the $pp$ scattering amplitude, and only performed the expansion in powers of $\eta$ when that amplitude is inserted in the expression for the asymmetry.

\subsection{Corrections proportional to the effective range}
\label{sec-reneq0}

Here we discuss corrections to the leading-order result of Eq.~(\ref{eq:ALppLO}). Since we only examine the form of the NLO correction, and do not compare to experimental data, we consider $nn$ scattering. The arguments are similar for $pp$ scattering and $np$ scattering.
Only operators with the same space-spin structure as the leading-order ones of Eq.~(\ref{Lag:RPS}) are necessary for this analysis. Other space-spin structures, e.g,  mixing between $P$- and $D$-waves, have the same number of derivatives as the operators we will consider in this section, but the resulting amplitudes are not enhanced by the strong $S$-wave rescattering. (E.g, effects of $P$-$D$ mixing do not enter until $O(Q^3)$: three orders beyond leading.)

The result in  Eq.~(\ref{eq:ALnn})  is actually somewhat deceptive. The use of the dibaryon formalism in the strong Lagrangian seems to imply that the physics of the effective range has been included to all orders in Eq.~(\ref{eq:ALnn}).  This is not the case because the dibaryon formalism was not used in the weak Lagrangian.  Either scaling for the effective range (as $Q^{-1}$ or $Q^0$) will lead to consistent results---but only  if the choice is used uniformly in all aspects of the calculation.

In particular, demanding that 
\begin{equation}
{\partial A_L^{nn} \over \partial \mu} = 0 
\label{eq:mudep}
\end{equation} implies that the ${\cal A}_{nn}$ of Eqs.~(\ref{eq:A1}) and (\ref{eq:A1b}) becomes energy dependent:
\begin{equation}
{\cal A}_{nn} \sim {1 \over {1 \over a}-{1 \over 2} r p^2 -\mu} \ \ ,
\label{eq:Anndib}
\end{equation}
where we have dropped the partial-wave specification on $a$ and $r$.
To obtain consistent results for the case considered here, where the effective range is natural ($\sim Q^0$), we must expand Eq.~(\ref{eq:Anndib}) in powers of $r$. This allows us to estimate 
the impact of corrections proportional to $r$ in the weak-interaction piece of the Lagrangian. It yields:
\begin{equation}
{\cal A}_{nn} \sim {1 \over {1 \over a} - \mu} \left(1 + 
{1 \over 2}{r p^2 \over {1 \over a} - \mu} + \cdots\right) \ \ .
\end{equation}
Writing this as
\begin{equation}
{\cal A}_{nn} = {\cal A}_{nn}^{LO} + p^2 \ {\cal A}_{nn}^{NLO} + \cdots,
\end{equation}
makes it clear that there must be corrections to the leading-order weak-interaction Lagrangian that have the same space-spin structure, but are proportional to the square of the momentum (equivalently, the energy) of the $NN$ collision. Each term will be accompanied by its own low-energy constant. These corrections to the
${\cal A}_{nn}$ of Eqs.~(\ref{eq:A1}) and (\ref{eq:A1b}) are suppressed by one power of the small parameter $Q$. They are missing from the result (\ref{eq:nnampl}), which includes only the leading-order part of ${\cal A}_{nn}$, and the effect of strong rescattering. Neglecting NLO contributions to the weak Lagrangian results in an inconsistent calculation as soon as $r \neq 0$.

While ${\cal A}_{nn}^{LO}$ recaptures the scaling of ${\cal C}_0$ (Eq. (\ref{eq:C0})), as expected,  ${\cal A}_{nn}^{NLO}$ runs like
${\cal C}_2$, the NLO strong coefficient (see, for example, Eq. (2.26) of Ref.~\cite{KSW98B}): 
\begin{equation}
{{\cal A}_{nn}^{NLO} \over {\cal A}_{nn}^{LO}} \sim {1 \over 2}{r \over {1 \over a} - \mu} \ \ .
\label{eq:NLOAscaling}
\end{equation}

The necessity for the weak Lagrangian to have an $O(Q)$ piece with coefficients scaling according to Eq.~(\ref{eq:NLOAscaling}) can also be derived by considering the $\mu$-invariance of the NLO amplitude for PV $NN$ scattering in a strictly perturbative calculation in powers of $Q$. Conversely, were we to use a weak Lagrangian expressed using dibaryon fields for the $S$-channels the scaling (\ref{eq:Anndib}) would emerge automatically. In either case, in order to maintain $\mu$-independent results the counting of $r$ must remain consistent between ${\cal L}_{weak}$ and 
${\cal L}_{strong}$. Both Lagrangians contain higher-order terms that are proportional to $r$, and the scaling of these contributions with $\mu$ is correlated. 
If effects proportional to $r$ are resummed using a dibaryon formalism consistently in both weak and strong Lagrangians, naive dimensional analysis suggests that additional corrections in $\mathcal{L}_{PV}$, which are related to additional parameters, are suppressed by two powers of Q.

\section{Comparison with experiment}

\label{sec-experiment}

In order to completely specify the leading-order PV $NN$ Lagrangian in this EFT the coefficients of the five dimension-7 operators must be determined. 
The only way to do this in a model-independent fashion is to fit them to experiment. If the experiments are at low enough energy the corrections from
higher-dimensional operators that encode other partial-wave transitions, as well as energy-dependence of the S-P transitions, will presumably be small.
In practice a higher-order analysis, together with a variety of different measurements, will have to be employed to 
see if the EFT is complete and consistent.

Here we pursue only a leading-order analysis of the two most recent low-energy measurements of the longitudinal asymmetry in $pp$ scattering. These yielded~\cite{Eversheim:1991tg}
 \begin{equation} 
A_L^{\vec p p}(E=13.6 \ {\rm MeV}) = (-0.93 \pm 0.21) \times 10^{-7}
\label{eq:Eversheim}
\end{equation}
and~ \cite{Kistryn:1987tq}
\begin{equation}
A_L^{\vec p p}(E=45 \ {\rm MeV}) = (-1.50 \pm 0.22) \times 10^{-7}
\label{eq:Kistryn}
\end{equation}
in the angular range $23^o <\theta_{lab} < 52^o$.

Using (\ref{eq:ALppfinal}) together with the lower-energy number (\ref{eq:Eversheim}) yields:
\begin{equation}
{\cal A}_{pp}(\mu=m_\pi)=(1.3 \pm 0.3) \times 10^{-14}~{\rm MeV}^{-3}.
\end{equation}
(Here and below the errors are only experimental, and do not include the uncertainty due to higher-order corrections.)  
For the $\mu$-independent ratio this gives:
\begin{equation}
\frac{{\cal A}_{pp}}{{\cal C}^{^1S_0}_0}=(-1.5 \pm 0.3) \times 10^{-10}~{\rm MeV}^{-1}.
\label{eq:muindLO}
\end{equation}

At this value of $p$ the Coulomb parameter $\eta=0.043$. The correction proportional to  $\eta$ in Eq.~(\ref{eq:ALppfinal}) also includes a factor of $1/(ap) \approx -0.13$. The Coulomb correction is only 3 percent, smaller than the uncertainties in the measurement and higher-order effects in \eftnopi.

Equation (\ref{eq:muindLO}) may be used to predict the scattering asymmetry at the higher energy of 45 MeV, yielding:
\begin{equation}
A_L^{\vec p p}(E=45 \ {\rm MeV}) =(-1.7 \pm 0.4) \times 10^{-7}.
\label{eq:LOpred}
\end{equation}

The two extant low-energy data are thus consistent with a leading-order \eftnopi~analysis within their combined uncertainties. This is really nothing more than the statement that at these energies the asymmetry is scaling with the center-of-mass momentum---as already observed in Ref.~\cite{Zhu:2004vw}. 

It should, however, be noted that the center-of-mass momentum for the second experiment is already larger than $m_\pi$. Sub-leading corrections could therefore be large. A crude estimate of these effects can be obtained by using Eq.~(\ref{eq:ALpp}), which includes the effects proportional to $r$ (but see also Sec.~\ref{sec-reneq0}) due to strong rescattering. This yields (with $r^{^1S_0}=2.73$ fm):
\begin{equation}
\frac{{\cal A}_{pp}(\mu=m_\pi)}{{\cal C}^{^1S_0}_0}=(-1.1 \pm 0.3) \times 10^{-10}~{\rm MeV}^{-1}.
\end{equation}
The shift of $\sim$ 30\% with respect to the leading-order value (\ref{eq:muindLO}) is entirely consistent with the expansion parameter of \eftnopi. The prediction for the higher-energy datum is now 
\begin{equation}
A_L^{\vec p p}(E=45 {\rm MeV}) =(-2.6 \pm 0.6) \times 10^{-7}.
\label{eq:NLOhigh}
\end{equation}
In this case the shift is more than 50\% of the leading-order value (\ref{eq:LOpred}), suggesting that the point at 45 MeV is indeed too high for profitable application of \eftnopi. The large (partial) $O(Q)$ correction computed here suggests that we can anticipate significant additional corrections at next-to-next-to-leading order. Given the presence of these corrections, as well as the experimental error, there is no real tension between (\ref{eq:NLOhigh}) and (\ref{eq:Kistryn}). The large NLO correction would, though, seem to imply that the agreement between the datum of Ref.~\cite{Kistryn:1987tq} and the LO prediction (\ref{eq:LOpred}) is fortuitous.

Measurements of the neutron's spin rotation as it passes through parahydrogen have been proposed,
e.g., in Ref.~\cite{Markoff:2005dm}. 
The hope here is to extract
the longitudinal analyzing power of $\vec n + p $  
scattering.  The thermal energies at which these experiments take
place are ideal for \eftnopi.  The leading-order
pionless EFT prediction is given in Eq.~(\ref{eq:ALnp}) and depends upon the coefficients $\aR$, $\bR$, $\dR$, and
$\eR$, so once low-energy data is available constraints on \eftnopi~coefficients will result.

\section{Conclusion and Outlook}

\label{sec-conclusion}

We have presented the leading-order low-energy prediction for $nn$, $pp$, and $np$ longitudinal asymmetries.  They depend on five different parameters, but one asymmetry measurement each in $\vec{n} + n$ and $\vec{p} + p$, as well as two at different energies in $\vec{n} + p$, would allow the extraction of four of the five parameters.  We determined, in agreement with the findings of earlier authors, that the Coulomb corrections to $pp$ scattering are not significant at leading order for the energies at which these measurements are made.  Finally, we showed that when the effective range is taken to scale as $Q^0$, the running of the leading order weak interaction coefficients mimics that of ${\cal C}_0$ of the strong interaction, while the running of the next-to-leading-order coefficients is expected to mimic that of ${\cal C}_2$ of the strong interaction.  This is a simple consequence of the mixing of the $S$-wave side of the parity violating operators with the bubble-sum enhancement of strong $S$-wave scattering.  

Our \eftnopi~calculations presented here, which implement a systematic power counting scheme, show the consistency of the Danilov hypothesis that (at least for energies $< m_\pi^2/M$) the dominant energy dependence in parity-violating $NN$ observables arises from the large $NN$ scattering lengths in the parity-conserving sector. Effects of energy (or momentum) dependence in the parity-violating $NN$ interaction, as well as those due to PV mixing between other partial waves, constitute higher-order effects in \eftnopi~which are accompanied by additional unknown parameters.

At leading order in $\eftnopi$ there are only five independent PV $NN$ operators~\cite{Girlanda:2008ts}. This means that five independent measurements will serve to pin down the leading-order Lagrangian.
Equation (\ref{Lag:RPS}) is one way to write the five terms of the LO PV Lagrangian in \eftnopi. It is equivalent to the previously published form (\ref{Lag:Gir}), with the matching computed in detail in the subsequent Appendix. The Lagrangian (\ref{Lag:RPS}) has the advantage of being written in an operator basis where each coefficient contributes to one and only one partial-wave transition.

For instance, the longitudinal asymmetry in $pp$ scattering probes the coefficients $\bR$, $\cR$, and $\dR$. Our formulae (\ref{eq:A2}) and (\ref{eq:ALpp}) encode the specific combination in which the coefficients associated with different isospin transitions appear. Existing experimental data can be used to extract this combination---admittedly with large error bars. A lower-energy $pp$ experiment with high precision would be a useful development. 

Meanwhile the partial-wave transitions $\aR$, $\bR$, $\dR$, and $\eR$ are probed in the $np$ longitudinal asymmetry. In principle a detailed study of the energy dependence of this asymmetry could allow the extraction of $\aR$ and the particular linear combination of the other three coefficients relevant for $np$ scattering.

Finally, the LO \eftnopi~prediction for the longitudinal analyzing power of $\vec n$ + $n$ scattering depends upon the coefficients $\bR$, $\cR$, and $\dR$, but in a different linear combination to that appearing in the prediction for $A_L^{pp}$. Given the difficulties inherent in such an experiment it seems more productive to focus on the asymmetry in $\vec{n} + d$ scattering at low energies (see also Ref.~\cite{Schiavilla:2008}) and perform the necessary three-body calculations for the interpretation of that asymmetry (see Ref.~\cite{BHvK98,BHvK00} for examples in the parity-conserving sector)
within the consistent \eftnopi~framework for parity-violating $NN$ scattering laid out here.

In any calculation of PV $NN$ observables it is important to treat the PV and PC $NN$ interactions consistently. In particular, care must be taken that operators and wave functions used in the same calculation are evaluated using compatible schemes and subtraction points. 
Use of the AV18 $NN$ potential to evaluate matrix elements of the short-range operators in Eq.~(\ref{Lag:Gir})~\cite{Liu:2006dm,Schiavilla:2008} represents a significant mismatch in this regard, and cannot be considered a systematic \eftnopi~calculation. 

In Sec.~\ref{sec-reneq0} we emphasized the importance of maintaining
a consistent power counting for both the weak and strong parts of
the Lagrangian.  A consistent calculation in \eftnopi can be carried
out assuming {\it either} that r scales as $Q^0$---in which case range
corrections are treated perturbatively---{\it or} that it scales as
$Q^{-1}$,
in which case a dibaryon formalism is necessary.  The most appropriate
choice should be revealed by seeing which (possibly higher-order) predictions provide  a better explanation of the data.

The PV coefficients for $NN$ asymmetries are presently 
experimentally underconstrained, so
it is necessary to use electromagnetic reactions in the $NN$ system to probe
additional linear combinations of the five PV parameters.
At lowest order in \eftnopi, both $\vec n p
 \rightarrow d \gamma$ and the anapole moment of the deuteron depend
 only upon a single coefficient  ($\eR$),  and so serve to
 disentangle this coefficient from the linear combination involved in
other $np$ processes. These have been computed in Ref.~\cite{Savage:2000iv}.
Measurements of the asymmetry in $\vec n p \rightarrow d \gamma$ are
presently consistent with
zero \cite{Cavaignac:1977uk,Alberi:1988fd}, but improvements by
an order of magnitude are expected \cite{Lauss:2006es,Seo}. 

A further constraint on the five PV parameters is potentially available from 
circularly polarized photon-deuteron breakup (or the inverse
reaction).  Experimentally, results are presently consistent with zero  \cite{Knyazkov:1984ke}.
The development of high intensity free electron lasers to
produce circularly polarized photons has led to proposals (e.g.,
Ref.~\cite{HIGS,JLAB}) to perform this measurement if the necessary
luminosity can be achieved. The PV parameters involved in the LO \eftnopi
prediction are $\aR$, $\bR$, and $\dR$.  This system has been discussed in Ref.~\cite{Khriplovich:2000mb}.
A partial LO calculation has recently been reported~\cite{Hyun:2008hp}.

We have presented a model-independent set of operators and
coefficients with which low-energy PV observables can be described and
compared, emphasizing the utility of the partial-wave basis. Such a treatment conveys significant advantages in our efforts to
understand manifestations of parity violation in few-nucleon systems.
Further calculations and experiments which use the \eftnopi~framework to map
out the landscape of possible experiments in two-, three-, and four-body
systems that are pertinent to parity violation would be very useful.

\section*{Acknowledgments}
DRP gratefully acknowledges the hospitality of the Theoretical Physics group at the University of Manchester and the Center for the Subatomic Structure of Matter at the University of Adelaide during part of this work. MRS would like to thank the Theoretical Physics group at the University of Manchester and the Lattice and Effective Field Theory group at Duke University for their hospitality. RPS acknowledges the hospitality of Ohio University, where much of this work was performed. We are grateful for discussions with Pil-Neyo Seo on the status of experiments. We would like to thank L.~Tiator for help obtaining Ref.~\cite{Eversheim:1991tg}. We thank D.~Eversheim for making the most recent analysis of the $13.6$ MeV longitudinal p-p scattering asymmetry experiment publicly available.This research was supported by DOE grants DE-FG02-93ER40756 (DRP and MRS) and DE-FG02-05ER41368 (RPS), and by an Ohio University Glidden Visiting Professorship.

\section{Appendix}  In this appendix we discuss the matching between 
the Weinberg basis Lagrange density of Eq.~(\ref{Lag:PC})
and the partial wave basis Lagrange density of Eq.~(\ref{Lag:PCpartial}).  
One method for matching the coefficients in these two different bases
is to use Fierz rearrangement identities.  Another method is to use the
Lorentz structures with their nucleon spin and isospin indices explicit,
and employ orthogonality and completeness of the operators in order to 
isolate one set of operators in terms of the other.  The latter is easy to implement
using a Mathematica \cite{wolfram} code with the HighEnergyPhysics `FeynCalc` package 
\cite{Mertig:1990an} to perform SU(2) manipulations.  

As a simple example we consider the lowest order terms of the strong
interaction Lagrange density for the two nucleon system. 
A useful Fierz identity is
$$(\sigma^\mu)_{ij} (\sigma_\mu)_{kl} =
2 \epsilon_{ik} \epsilon_{jl} \ \ ,$$
where $\epsilon=i\sigma_2$ is the totally anti-symmetric Levi-Civita tensor,
$\mu$ is summed over ($\mu=0,1,2,3$), $\sigma^\mu=(1,\vec \sigma)$ and $i,j,k,$ and $l$
are summed over spin indices.  An identical
equation serves just as well for the $\tau$ matrices and the isospin
indices (we will use $a,b,c$, and $d$ for these).  Putting this together:

$$(\sigma^\mu)_{ij} (\sigma_\mu)_{kl} (\tau^\nu)_{ab} (\tau_\nu)_{cd}  =
(2 \epsilon_{ik} \epsilon_{jl})( 2 \epsilon_{ac} \epsilon_{bd})=
4  (\sigma_2)_{ik} (\sigma_2)_{jl} (\tau_2)_{ac} (\tau_2)_{bd} \ \ . $$

To obtain an equality involving the $C_S$ operator, $(\ND N) (\ND N)$,
we want delta
functions rather than $\sigma_2$'s and $\tau_2$'s on the right hand side,
so act on both sides with $(\sigma_2)_{i^\prime i} (\sigma_2)_{l l^\prime}
(\tau_2)_{a^\prime a} (\tau_2)_{d d^\prime}$ and obtain
\begin{eqnarray}\label{delta}
(\sigma_2 \sigma^\mu)_{i^\prime j} (\sigma_\mu \sigma_2)_{kl^\prime} 
(\tau_2 \tau^\nu)_{a^\prime b} (\tau_\nu \tau_2)_{cd^\prime}  =
4  \delta_{i^\prime k} \delta_{jl^\prime} \delta_{a^\prime c} 
\delta_{bd^\prime} \ \ . 
\end{eqnarray}
Contracting this with $N^\dagger_{kc} N_{i^\prime a^\prime}
N^\dagger_{l^\prime d^\prime} N_{jb}$ -- the nucleon operators with
their spin and isospin indices explicit -- and noticing that
diagonal terms with $N^T \sigma_2 \tau_2 N$ and $N^T \sigma_2 \sigma_i \tau_2
\tau_j N$ 
are disallowed by the Pauli principle,  we obtain the decomposition
of the operator associated with $C_S$ in terms of the operators in the
partial wave basis:
$$(\ND N) (\ND N) = -2
(N^T P_a(^1 \! S_0) N)^\dagger (N^T P_a(^1 \! S_0) N) -
2 (N^T P_i(^3 \! S_1) N)^\dagger (N^T P_i(^3 \! S_1) N) \ \ , $$
where  $a$ and  $i$ are now summed over (1,2,3) as in the projection
operators introduced in section~\ref{sec-Lag}.

To obtain the form $(\ND \sigma_i N) (\ND \sigma_i N)$, associated
with $C_T$, out of the
Fierz identity, we need to appropriately insert not only $\sigma_2$ and 
$\tau_2$ (to get to delta functions) but $\sigma_A$'s on the right-hand side as well. Contracting 
$$ (\sigma_A)_{mi^\prime} (\sigma_A)_{l^\prime n} \ND_{ma^\prime}
N_{kc} \ND_{jb} N_{n d^\prime} $$ 
on both sides of Eq.~(\ref{delta}) yields,
\begin{eqnarray}
4 (\ND \sigma_i N) (\ND \sigma_i N) = 
(N^* \sigma_i \sigma_2 \tau_2 \ND) (N^T \sigma_2 \sigma_i \tau_2 N) -
(N^* \sigma_i \sigma_2 \tau_2 \tau_j \ND) 
 (N^T \sigma_2 \sigma_i \tau_j \tau_2 N) - \nonumber \\
(N^* \sigma_i \sigma_2 \sigma_j \tau_2 \ND) 
 (N^T \sigma_j \sigma_2 \sigma_i \tau_2 N) + 
(N^* \sigma_i \sigma_2 \sigma_j \tau_2 \tau_k \ND) 
 (N^T \sigma_j \sigma_2 \sigma_i \tau_k \tau_2 N).
\end{eqnarray}
But this can be considerably simplified because
\begin{eqnarray}
(N^* \sigma_i \sigma_2 \tau_2 \tau_j \ND) 
 (N^T \sigma_2 \sigma_i \tau_j \tau_2 N) &=& 0 \ \ ; \nonumber \\
(N^* \sigma_i \sigma_2 \sigma_j \tau_2 \ND) 
 (N^T \sigma_j \sigma_2 \sigma_i \tau_2 N) &=& 2 
(N^* \sigma_i \sigma_2 \tau_2 \ND) (N^T \sigma_2 \sigma_i \tau_2 N) 
\ \ ;  \nonumber \\
(N^* \sigma_i \sigma_2 \sigma_j \tau_2 \tau_k \ND) 
 (N^T \sigma_j \sigma_2 \sigma_i \tau_k \tau_2 N) & =&  3 
(\ND \sigma_2 \tau_k \tau_2 \ND) (N \sigma_2 \tau_2 \tau_k  N),\nonumber\\
\end{eqnarray}
so that
\begin{eqnarray}
(\ND \sigma_i N) (\ND \sigma_i N) =  
-2 (N^T P_a(^1 \! S_0) N)^\dagger (N^T P_a(^1 \! S_0) N) +
6 (N^T P_i(^3 \! S_1) N)^\dagger (N^T P_i(^3 \! S_1) N),\nonumber\\
\end{eqnarray}
which completes the decomposition of the $C_S$ and $C_T$ operators
in terms of the ${\cal C}_0^{^3 \! S_1}$ and  ${\cal C}_0^{^1 \! S_0}$ 
operators. 

\begin{figure}
\includegraphics[width=6cm]{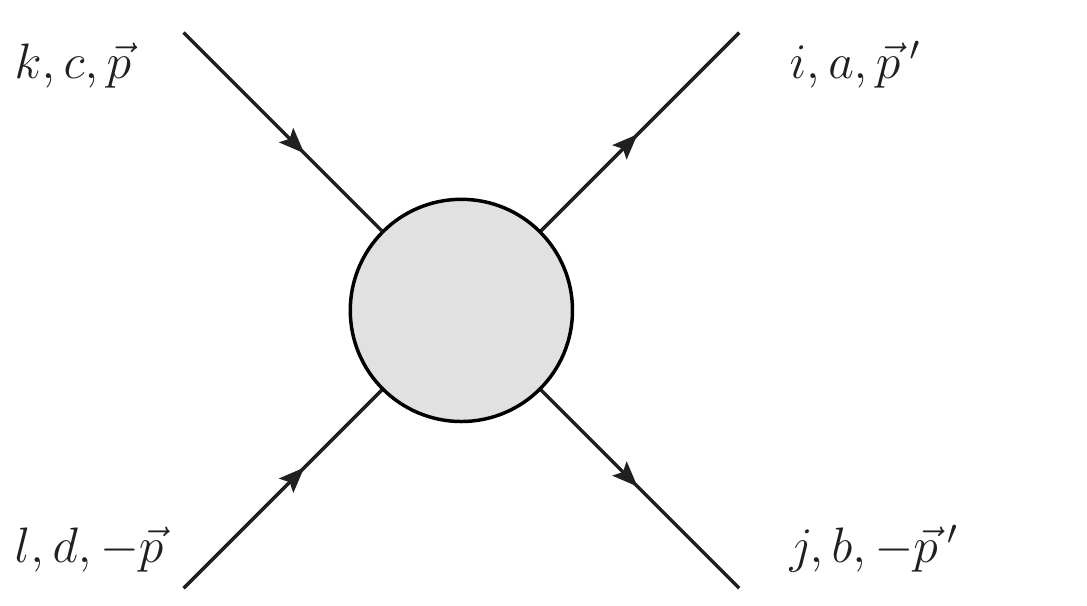}\caption{Assignment of spin indices ($i,j,k,l$), isospin indices ($a,b,c,d$), and momentum labels for purposes of matching the partial wave basis operators to the strong Weinberg and weak Girlanda operators. Figure created using JaxoDraw \cite{Binosi:2003yf}.
\label{Fig:match}}
\end{figure}

An easier procedure for obtaining one set of basis coefficients in
terms of the other is to use orthogonality and completeness of the operator
sets.  We will illustrate this using, again, the leading-order strong-interaction terms. 
Making the spin indices ($i,j...$) and isospin indices ($a,b...$) explicit
(referring to Fig.~\ref{Fig:match}) and including all possible nucleon assignments,
\begin{align}
- C_S \delta_{ik} \ \delta_{ac} \ \delta_{jl} \ \delta_{bd}  +
C_S  \delta_{jk} \ \delta_{bc} \ \delta_{il} \ \delta_{ad}  & - 
C_T  (\sigma_A)_{ik} \ \delta_{ac} \ (\sigma_A)_{jl} \ \delta_{bd} +
C_T \ (\sigma_A)_{jk} \ \delta_{bc} \ (\sigma_A)_{il} \ 
\delta_{ad}  \ = \notag \\
{1 \over 2} \ {\cal C}_0^{(^1 \! S_0)}   (\tau_A \tau_2)_{ab} \
(\sigma_2)_{ij} \ (\tau_2 \tau_A)_{cd} \ (\sigma_2)_{kl} &+
{1 \over 2} \ {\cal C}_0^{(^3 \! S_1)} \  (\tau_2)_{ab} \
(\sigma_B \sigma_2)_{ij} \ (\tau_2)_{cd} \ (\sigma_2 \sigma_B)_{kl} \ \ ,
\end{align}
where summation over $A,B=1,2,3$ is implied. 
Contracting both sides with the first structure, 
$\delta_{ik} \ \delta_{ac} \ \delta_{jl} \ \delta_{bd}$,
yields:$${\cal C}_0^{(^1 \! S_0)}+{\cal C}_0^{(^3 \! S_1)} =
2 C_S-2 C_T \ \ . $$
A second equation is found by contracting both sides with the third structure,
$(\sigma_A)_{ik} \ \delta_{ac} \ (\sigma_A)_{jl} \ \delta_{bd}$,
yielding:
$$3 {\cal C}_0^{(^1 \! S_0)}-{\cal C}_0^{(^3 \! S_1)} =
2 C_S-10  C_T \ \ . $$  Solving for the partial wave coefficients
yields the relationships given in Section~\ref{sec-Lag}.

Now consider the two weak-interaction bases from Eq.~(\ref{Lag:Gir})
and Eq.~(\ref{Lag:RPS}). Note that the operators in Eq.~(\ref{Lag:Gir}) are explicitly hermitian,
but not symmetric under interchange of outgoing (or incoming) particles.
On the other hand, the basis used in Eq.~(\ref{Lag:RPS}) is symmetric under interchange of
particles, but each is not its own hermitian conjugate.  This is important
to remember when comparing coefficients.

Even without Fierzing, inspection of the 
operators suggests that not all of the partial wave operators are
involved in the decomposition of, say, the operator associated
with $\CG_1$ in Eq.~(\ref{Lag:Gir}).  But no orthogonality need
be assumed because it is easy to verify.  
The starting point is (momenta from Fig.~\ref{Fig:match}):
{\allowdisplaybreaks 
\begin{align} 
&4 \aR (\sigma_A \sigma_2)_{ij} (\sigma_2)_{kl} 
 (\tau_2)_{ab} (\tau_2)_{cd} (2p_A) +
4 \aR (\sigma_2 \sigma_A)_{kl} (\sigma_2)_{ij} 
  (\tau_2)_{ab} (\tau_2)_{cd} (2p^\prime_A)  \notag \\
& +4 \bR (\sigma_2)_{ij} (\sigma_2 \sigma_A)_{kl}  (\tau_B \tau_2)_{ab}
(\tau_2 \tau_B)_{cd} (2p_A)  \nonumber \\
& +4 \bR (\sigma_2)_{kl} (\sigma_A \sigma_2)_{ij} 
   (\tau_B \tau_2)_{ab} (\tau_2 \tau_B)_{cd} (2p^\prime_A ) \nonumber \\
& +4 \cR (\sigma_2)_{ij} (\sigma_2 \sigma_A)_{kl} 
  (\tau_B \tau_2)_{ab} (\tau_2 \tau_C)_{cd} \epsilon_{3BC} (-2 i p_A) \nonumber \\
& +4 \cR (\sigma_2)_{kl} (\sigma_A \sigma_2)_{ij} 
  (\tau_2 \tau_B)_{cd} (\tau_C \tau_2)_{ab} \epsilon_{3BC} (-2 i p^\prime_A) \nonumber \\
& +4 \dR (\sigma_2)_{ij} (\sigma_2 \sigma_A)_{kl} (\tau_B \tau_2)_{ab}
 (\tau_2 \tau_C)_{cd} \mathcal{I}_{BC} (2 p_A)  \nonumber \\
& +4 \dR (\sigma_2)_{kl} (\sigma_A \sigma_2)_{ij} (\tau_2 \tau_B)_{cd}
  (\tau_C \tau_2)_{ab} \mathcal{I}_{BC} (2 p^\prime_A)  \nonumber \\
& +4 \eR (\sigma_A \sigma_2)_{ij} (\sigma_2 \sigma_C)_{kl} 
  (\tau_2)_{ab} (\tau_2 \tau_3)_{cd} \epsilon_{ABC}(-2 i p_B) \nonumber \\
& +4 \eR (\sigma_2 \sigma_A)_{kl} (\sigma_C \sigma_2)_{ij} 
 (\tau_2)_{cd} (\tau_3 \tau_2)_{ab} \epsilon_{ABC}
 (-2 i p^\prime_B) \nonumber \\
& = 2 \CG_1  ((\sigma_A)_{ik} \delta_{jl} - \delta_{ik} (\sigma_A)_{jl}) 
 (p+p^\prime)_A \delta_{ac} \delta_{bd}  \nonumber \\
& -2 \CG_1 ((\sigma_A)_{jk} \delta_{il} - \delta_{jk} (\sigma_A)_{il})
 (p-p^\prime)_A \delta_{bc} \delta_{ad} \nonumber \\
& -2 \tilde \CG_1 \epsilon_{ABC} (\sigma_A)_{ik} (\sigma_C)_{jl}
(ip^\prime-i p)_B \delta_{ac} \delta_{bd}  \nonumber \\
& +2 \tilde \CG_1 \epsilon_{ABC} (\sigma_A)_{jk} (\sigma_C)_{il}
(-ip^\prime-i p)_B \delta_{bc} \delta_{ad} \nonumber \\
& -2 \CG_2 \epsilon_{ABC} (\sigma_A)_{ik} (\sigma_C)_{jl} (i p^\prime - i p )_B
 ((\tau_3)_{ac} \delta_{bd} +\delta_{ac} (\tau_3)_{bd})  \nonumber \\
& +2 \CG_2 \epsilon_{ABC} (\sigma_A)_{jk} (\sigma_C)_{il} (-i p^\prime - i p )_B
 ((\tau_3)_{bc} \delta_{ad} +\delta_{bc} (\tau_3)_{ad}) \nonumber \\
& -2 \tilde \CG_5 \epsilon_{ABC} (\sigma_A)_{ik} (i p^\prime - i p)_B 
(\sigma_C)_{jl} {\mathcal I}_{DE} (\tau_D)_{ac} (\tau_E)_{bd} \nonumber \\
& +2 \tilde \CG_5 \epsilon_{ABC} (\sigma_A)_{jk} (-i p^\prime - i p)_B 
(\sigma_C)_{il} {\mathcal I}_{DE} (\tau_D)_{bc} (\tau_E)_{ad} \nonumber \\
& +2 \CG_6 \delta_{ik} (\sigma_A)_{jl} (-i p^\prime + i p)_A \epsilon_{BC3}
 (\tau_B)_{ac} (\tau_C)_{bd}  \nonumber \\
& -2 \CG_6 \delta_{jk} (\sigma_A)_{il} (i p^\prime + i p)_A \epsilon_{BC3}
 (\tau_B)_{bc} (\tau_C)_{ad}  \ \ .
\end{align}}
Again, $i,j,...$ are the spin indices and $a,b,...$ the isospin indices.
The factor of 4 on the left hand side comes 
from all possible nucleon assignments.  The coefficients
are real.  Contracting both sides with one operator (in, for example, the
partial wave basis) at a time yields a set of equations involving
coefficients only.  For example, contracting with $
(\sigma_A \sigma_2)_{ij} (\sigma_2)_{kl} 
 (\tau_2)_{ab} (\tau_2)_{cd} (2p_A) $ yields 
$\aR={1 \over 4} (\CG_1-\tilde \CG_1)$. 
Provided both sets of operators are minimal and
complete (at the order desired), there will be a unique solution. 
The procedure is systematized using Mathematica \cite{wolfram} and the results are
provided in Section~\ref{sec-Lag}.

\end{document}